\documentclass[aps,prb,amsmath,reprint,showpacs,twocolumn,amssymb,titlepage,superscriptaddress]{revtex4-1}

\usepackage{graphicx}
\usepackage{nicefrac}
\usepackage{amsfonts}

\usepackage{amssymb}
\usepackage{amsmath} 

\usepackage{subfigure}
\usepackage{multirow} 
\usepackage{tabularx} 
\usepackage{array}
\usepackage{adjustbox}
 \DeclareGraphicsExtensions{.pdf,.jpeg,.png,.jpg} 

\usepackage{units}

\usepackage{tensor} 
\usepackage{braket}

\usepackage{bm}
\usepackage{hyperref}

\begin{document}
\title{Low-temperature acanthite-like phase of Cu$_{2}$S: A first-principles study on electronic and transport properties}
\author{Ho Ngoc Nam}\thanks{Email: honam@mat.eng.osaka-u.ac.jp}
\address{Department of Precision Science and Technology, Graduate School of Engineering, Osaka University, 2-1 Yamadaoka, Suita, Osaka 565-0871, Japan}
\address{Division of Materials and Manufacturing Science, Graduate School of Engineering, Osaka University, 2-1 Yamadaoka, Suita, Osaka 565-0871, Japan}
\author{Katsuhiro Suzuki}
\address{Division of Materials and Manufacturing Science, Graduate School of Engineering, Osaka University, 2-1 Yamadaoka, Suita, Osaka 565-0871, Japan}
\author{Tien Quang Nguyen}
\address{Division of Materials and Manufacturing Science, Graduate School of Engineering, Osaka University, 2-1 Yamadaoka, Suita, Osaka 565-0871, Japan}
\author{Akira Masago}
\address{Center for Spintronics Research Network, Graduate School of Engineering Science, Osaka University, Toyonaka, Osaka 560-8531, Japan}
\author{Hikari Shinya}
\address{Center for Spintronics Research Network, Graduate School of Engineering Science, Osaka University, Toyonaka, Osaka 560-8531, Japan}
\address{Research Institute of Electrical Communication, Tohoku University, Sendai 980-8577, Japan}
\address{Center for Spintronics Research Network, Tohoku University, Sendai, Miyagi 980-8577, Japan}
\author{Tetsuya Fukushima}
\address{Center for Spintronics Research Network, Graduate School of Engineering Science, Osaka University, Toyonaka, Osaka 560-8531, Japan}
\address{Institute of Solid State Physics, The University of Tokyo, 5-1-5 Kashiwanoha, Kashiwa, Chiba, 277-8581, Japan}
\author{Kazunori Sato}
\address{Division of Materials and Manufacturing Science, Graduate School of Engineering, Osaka University, 2-1 Yamadaoka, Suita, Osaka 565-0871, Japan}
\address{Center for Spintronics Research Network, Graduate School of Engineering Science, Osaka University, Toyonaka, Osaka 560-8531, Japan}
\date{\today}
%
%
%
%
%
\begin{abstract}
The mobility and disorder in the lattice of Cu atoms as liquid-like behavior is an important characteristic affecting the thermoelectric properties of Cu$_{2}$S. In this study, using a theoretical model called acanthite-like structure for Cu$_{2}$S at a low-temperature range, we systematically investigate the electronic structure, intrinsic defect formation, and transport properties by first-principles calculations. Thereby, previous experimental reports on the indirect bandgap nature of Cu$_{2}$S were confirmed in this work with an energy gap of about 0.9--0.95 eV. As a result, the optical absorption coefficient estimated from this model also gives a potential value of $\alpha > 10^{4}$ cm$^{-1}$ in the visible spectrum range. According to the bonding analysis and formation energy aspect, Cu vacancy is the most preferred defect to form in Cu$_{2}$S, which primarily affects the conductive behavior as a $p$-type, as experimentally observed. Finally, the transport properties of Cu$_{2}$S system were successfully reproduced using an electron-phonon scattering method, highlighting the important role of relaxation time prediction in conductivity estimation instead of regarding it as a constant.
\end{abstract}
%
%
%
%
%
%
\maketitle
\section{Introduction}
Since its inception, thermoelectric (TE) technology has been proving its robust potential for harvesting waste heat and converting it into electricity. \cite{r1,r2} This green solution turned out to be extremely useful in current circumstances, especially when fossil fuel resources are running out and the challenges of climate change are constantly increasing.\cite{r3} Thanks to these practical impetuses, many TE material generations have been explored with impressive performances.\cite{r4,r5} One of the most popular TE material paradigms called the ``\emph{phonon-glass electron-crystal}" (PGEC), was proposed in 1995.\cite{r6,r7} Interestingly, most of the state-of-art TE materials so far are in the PGEC paradigm with crystalline solid form. In nature, the thermal conductivity of liquids is known to be worse than that of solids, and this concept is a hint we can exploit to renovate the PGEC paradigm. Since 2012, such a TE material paradigm has been around, called a ``\emph{phonon-liquid electron-crystal}" (PLEC), which takes advantage of the liquid-like behavior of superionic conductors in several materials to minimize thermal conductivity.\cite{r8,r9} In particular, transition metal-chalcogenide compounds such as (Ag,Cu)$_{2}$(S,Se,Te) group are typical materials for this concept with many fascinating properties and high TE performance.\cite{r10,r11,r12,r13}

Copper sulfide or Cu$_{2-x}$S is inherently not only a well-known semiconductor for applications in photovoltaic solar cells,\cite{r14,r15} but recently it has also been recognized as a promising candidate for TE applications.\cite{r16} Although Cu$_{2}$S has been studied since the late 1940s, the thorough understanding of the crystal structure and electronic properties still has some ambiguity and controversy.\cite{r17,r18,r19} Referring to the crystal structure issue first, Cu$_{2-x}$S exists in many crystallographically distinct phases depending on the Cu content, such as chalcocite (Cu$_{2}$S), djurleite (Cu$_{1.96}$S), digenite (Cu$_{1.8}$S), anilite (Cu$_{1.75}$S). In which, stoichiometric compound Cu$_{2}$S has three temperature-dependent phases including the $\gamma$-phase (low-chalcocite or \emph{L}-chalc. for temperatures below 378 K), the $\beta$-phase (high-chalcocite or \emph{H}-chalc. in range of 378 K to 698 K), and the $\alpha$-phase (above 698 K).\cite{r20}

However, it seems not to be that simple. The sensitivity to the temperature of Cu atoms makes them really mobile and disorderly, which is considered as liquid-like behavior. As a result, locating atoms becomes difficult and confuses us in the study of their properties. For instance, in previous studies,\cite{r21,r22} the authors investigated the electronic structure of Cu$_{2}$S based on several artificial models for high-temperature phase. Despite also considering the available low-temperature phase \emph{L}-chalc. determined from the experiment,\cite{r19} the results in these models still do not reproduce the bandgap energy or reveal the indirect nature as observed in the experimental report.\cite{r23,r24} The problem is, \emph{L}-chalc. phase found to have a large pseudo-orthorhombic structure including 96 molecular units (288 atoms) where each Cu atom has a unique site. Obviously, such a large number of atoms along with the low symmetry of structure cause density functional theory (DFT) based investigations more challenging. To simplify this hobble, a previous study tried to reduce the number of atoms by approximately dividing it into two identical monoclinic cells (144 atoms per cell).\cite{r25} Nevertheless, this is still a modified structure with a large number of atoms. Secondly, it is not just about the issue of computational cost. The disorder of the Cu atoms and a large supercell size can lead to the folding of $k$-points at the Brillouin-zone edge on the $\Gamma$-point. Consequently, it can make the bandgap direct while the evidence from the optical data shows that the nature of the bandgap should be indirect. Recently, in a systematic structural investigation,\cite{r26} the author used 15 possible crystal structures for Cu$_{2}$S to estimate the most stable one based on cohesive energy. Accordingly, a new phase derived from a similar low-temperature phase of Ag$_{2}$S called the acanthite-like phase, which reveals that it is the most stable structure at 0 K of Cu$_{2}$S theoretically. Intriguingly, this acanthite-like structure is pretty simple. The arrangement of Cu and S atoms forms layered chains with a zig-zag shape where Cu atoms occupy only two order positions, tetrahedral and octahedral sites. Since there is not much structural difference (both are monoclinic), \emph{L}-chalc. phase and acanthite-like phase are reported to be pretty similar in the electronic structure except for direct/indirect nature. Therefore, it would be better to have insight into the physical properties of Cu$_{2}$S in the acanthite-like structure, which is theoretically the most stable one. 

In this work, we systematically investigate the electronic properties, the intrinsic point defect formation, and transport properties of the acanthite-like phase of Cu$_{2}$S by using first-principle calculations. In section A, the electronic structure of Cu$_{2}$S is considered using an acanthite-like model. In section B, the formation of point defects and diffusion behavior are also discussed as a part of electronic properties. Finally, in section C, the transport properties of the acanthite-like model based on the electron-phonon scattering mechanism and the rationality of this model for TE designing purposes are the main parts we focused on this study.
%
%
%
%
%
%
\section{Computational methods}
\subsection{Electronic properties}
Our DFT calculations are mainly carried out using VASP code.\cite{r27} The projector-augmented wave (PAW) approach\cite{r28} is used with generalized gradient approximation (GGA) in the form of Perdue-Burke-Ernzerhof (PBE).\cite{r29} Here, the \emph{3d$^{10}$} and \emph{4s$^{1}$} electrons of Cu, \emph{3s$^{2}$} and \emph{3p$^{4}$} electrons of S are treated as valence states. The wave functions are expanded in a plane-wave basis set with cut-off energy of 400 eV.  A width of 0.05 eV of Gaussian smearing has been used this work. All calculations were converged until the residual atomic force becomes smaller than 10$^{-2}$ eV/Å.
Since the DFT method is well-known for underestimating bandgap energy, the rotationally invariant DFT+$U$ method with an effective Hubbard parameter \emph{U} of 7 eV,\cite{r30,r31} was applied on \emph{d}-orbital of Cu to handle on-site Coulomb interaction. Besides, we also employed the hybrid functional proposed by Heyd, Scuseria, and Ernzerhof (HSE06)\cite{r32} in several cases related to band structure calculations to compare with the results from the DFT+$U$ method. The BZ was sampled using the Monkhorst-Pack \emph{k}-mesh of 11x9x9 for PBE functional, while a 8x4x4 mesh was used for HSE06 functional. For defect formation calculations, a 2x2x2 supercell with a corresponding 3x3x3 \emph{k}-meshes is used. The migration pathways of Cu vacancy were discussed as well based on possible minimum energy pathways (MEP) between the adjacent sites using the nudged elastic band (NEB) algorithm.\cite{r33}

In the framework of band structure calculations, the linear optical properties can be obtained from the frequency-dependent complex dielectric function\cite{r34}:
\begin{equation}
\varepsilon (\omega )=\varepsilon _{1}(\omega )+i\varepsilon _{2}(\omega ),
\end{equation}
where $\varepsilon _{1}(\omega )$ and $\varepsilon _{2}(\omega )$ are the real and imaginary parts of the dielectric function, respectively; $\omega$ is the photon frequency. Consequently, the absorption coefficient $\alpha (\omega )$ is derived from $\varepsilon _{1}(\omega )$ and $\varepsilon _{2}(\omega )$ as following:
\begin{equation}
\alpha (\omega )=\frac{\sqrt{2}\omega }{c}\left ( \sqrt{\varepsilon _{1}^{2}+\varepsilon _{2}^{2}} -\varepsilon _{1} \right )^{\frac{1}{2}}.
\end{equation}
In addition, the electric transition dipole moment (TDM) based on dipole transition matrix elements $\textup{P}_{a \to b }$ between two states is defined as\cite{r35}
\begin{equation}
\textup{P}_{a \to b }=\left \langle \psi_{b}\left | \mathbf{r} \right |\psi _{a} \right \rangle=\frac{i\hbar}{(E_{b}-E_{a})m}\left \langle  \psi_{b}\left | \mathbf{p} \right |\psi _{a}\right \rangle,
\end{equation}
where $\psi_{a}$ and $\psi_{b}$ are eigenstates corresponding to energy $E_{a}$ and $E_{b}$, $m$ is the electron mass. It should be noted that for a more accurate band structure description, the meticulous estimation may necessitate calculations such as the GW-BSE method.\cite{r36} Only the predictions of the GGA+$U$ and HSE06 functionals are considered here.
\subsection{Intrinsic point defects formation}
The defect formation energy $E_{form}(D,q)$ at charge state $q$ of defect $D$ as a function of Fermi energy can be defined as following\cite{r37}
\begin{equation}
\begin{split}
&E_{form}(D,q) = E^{tot}_{D,q} - E^{tot}_{bulk} + \sum{n(i)\mu_{i}} \\ 
& + q(E_{\textup{VBM}} +\Delta{E}_{\textup{Fermi}}),
\end{split}
\end{equation}
where $E^{tot}_{D,q}$ is the total energy of the defect system, $E^{tot}_{bulk}$ is the total energy of the bulk system, $n$ is the number of impurity atoms ($n>0$ for doped atoms and $n<0$ for removed atoms), $\mu_{\textup{Cu}}$ and $\mu_{\textup{S}}$ are chemical potentials of Cu and S, respectively. $E_{\textup{VBM}}$ is referenced energy related to the valence band maximum (VBM) while $\Delta{E}_{\textup{Fermi}}$ is Fermi energy relative to VBM. We also used simply core potential correction as a correction term for Eq. (4). Then, the chemical potentials can be calculated by the thermodynamic equilibrium conditions as:
\begin{equation}
2\mu _{\textup{Cu}}+\mu _{\textup{S}}=\mu _{\textup{Cu}_{2}\textup{S}(bulk)}
\end{equation}
\begin{equation}
\mu _{\textup{Cu}}\leq \mu _{\textup{Cu}}^{bulk}+\Delta \mu _{\textup{Cu}}
\end{equation}
\begin{equation}
\mu _{\textup{S}}\leq \mu _{\textup{S}}^{bulk}+\Delta \mu _{\textup{S}}
\end{equation}
where $\mu_{\textup{Cu}_{2}\textup{S}(bulk)}$ is the chemical potential of the bulk system with $\mu_{\textup{Cu}_{2}\textup{S}(bulk)}$ = -12.01 eV. The fcc Cu bulk and S$_{8}$ molecule are used for ascertaining $\mu_{\textup{Cu}}$ and $\mu_{\textup{S}}$, respectively. Here, we basically consider two main crystal growth conditions, namely Cu-rich condition ($\Delta \mu _{\textup{Cu}}$= 0) with $\mu _{\textup{Cu}}=\mu _{\textup{Cu}}^{bulk}$= -3.69 eV and Cu-poor condition ($\Delta \mu _{\textup{S}}$= 0) with $\mu _{\textup{S}}=\mu _{\textup{S}}^{bulk}$= -4.12 eV.
\subsection{Transport properties: electron-phonon coupling}
For transport properties, the effect of electron-phonon coupling is investigated using Quantum Espresso code.\cite{r38} Here, we used a plane-wave basis set with kinetic energy cutoffs of 60 and 600 Ry for wave functions and charge density, respectively. Besides, the uniform 12x12x12 $\Gamma$-centered $k$-point and 3x3x3 $q$-point grids are used for calculations. Then, the general transport parameters of the system are calculated using BoltzTraP code\cite{r39} to solve the semiclassical Boltzmann transport equation within the relaxation-time approximation. The expressions for electrical conductivity ($\sigma$), Seebeck coefficient ($S$), electronic thermal conductivity ($\kappa ^{e}$) are the following:\cite{r40}
\begin{figure*}[htbp]
\centerline{\includegraphics{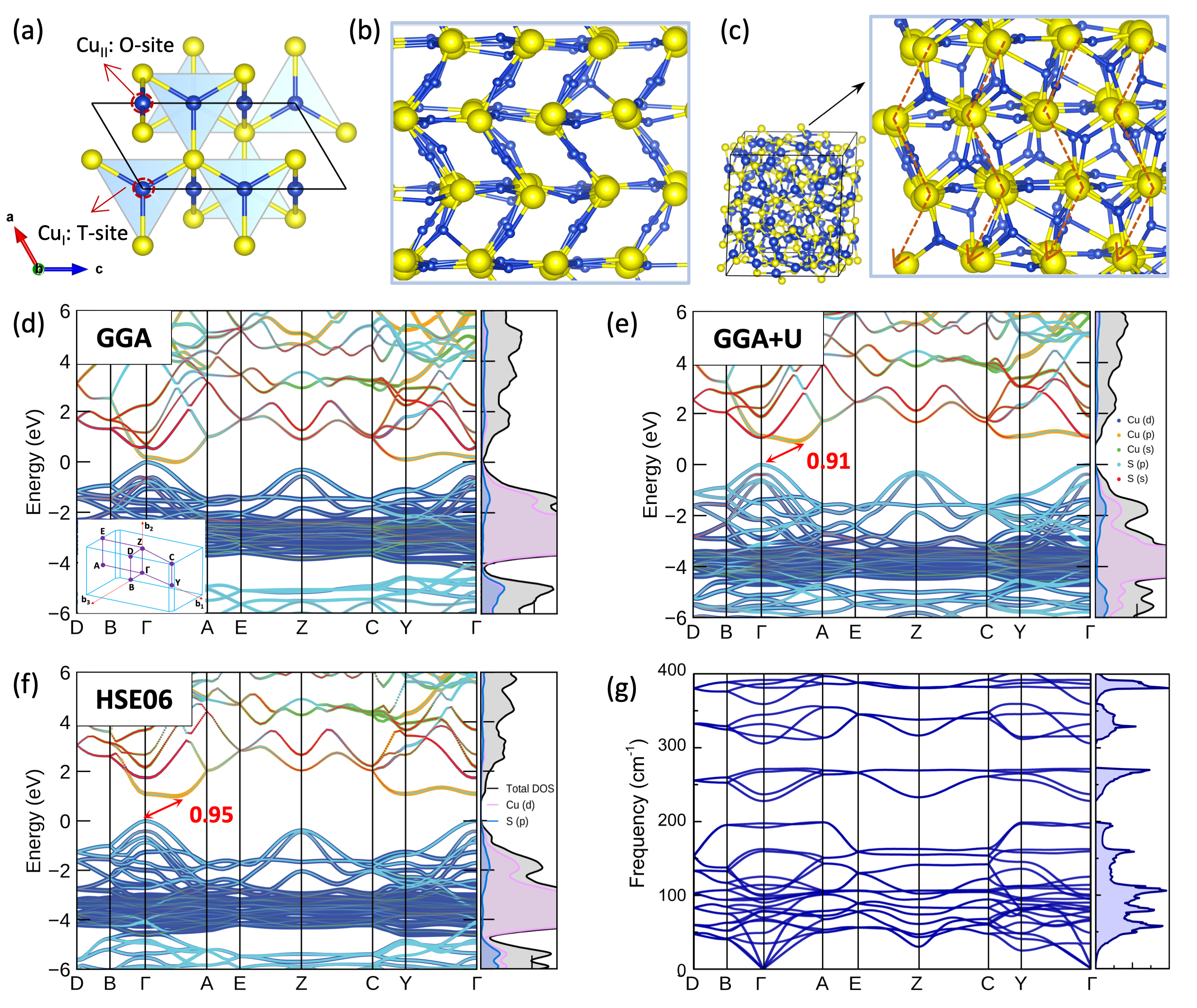}}
\caption{(a) Crystal structure of Cu$_{2}$S acanthite-like phase (Cu atom: blue, S atom: yellow) where Cu occupied two significant positions, tetrahedral-site (Cu$_\textup{I}$) and octahedral-site (Cu$_\textup{II}$). (b) The displacement of Cu atoms in the acanthite-like phase via Ab-initio molecular dynamics simulation at 300 K shows similar behavior to the one in the low-chalcocite phase observed in experiments (c), where the S sub-lattice is nearly immobile with the zig-zag arrangement. (d-f) Electronic band structure, Brillouin zone, and DOS of Cu$_{2}$S acanthite-like phase by different functionals (Fermi energy is referred to the top of valence band); (g) Phonon dispersion and phonon DOS of acanthite-like phase.}
\label{Fig:pic1}
\end{figure*}

\begin{equation}
\sigma _{\alpha \beta }(\mu ,T)=K_{\alpha \beta }^{(0)}
\end{equation}
\begin{equation}
S _{\alpha \beta }(\mu ,T)=k_{B}\sum _{i}(K^{(0)^{-1}})_{i\alpha }K_{i\beta }^{(1)}
\end{equation}
\begin{equation}
\kappa ^{e} _{\alpha \beta }(\mu ,T)=k^{^{2}}_{B}T\left [K^{(2)}_{\alpha \beta }-\sum _{ij}K_{\alpha i }^{(1)}(K^{(0)^{-1}})_{ij}K_{j\beta }^{(1)}  \right ] 
\end{equation}
where $\alpha$, $\beta$, $i$, $j$ are Cartesian componentsm, $\mu$ is the chemical potential, $T$ is the absolute temperature, $k_{B}$ is the Boltzmann constant. Here, $K ^{(p)} _{\alpha \beta }$ is the \emph{p}-th order electronic transport coefficient, which is defined as
\begin{equation}
\begin{split}
&K ^{(p)} _{\alpha \beta }(\mu ,T)=\frac{g_{s}e^{2-p}}{(2\pi )^{3}(k_{B}T)^{p+1}}\sum _{n}\int _{BZ}d\mathbf{k}v_{n\mathbf{k}\alpha }v_{n\mathbf{k}\beta } \\
&\times \tau _{n\mathbf{k}}(\mu ,T)I^{(p)}(\epsilon _{n\mathbf{k}},\mu ,T)
\end{split}
\end{equation}
with $g_{s}$ is spin degeneracy, $\mathbf{k}$ is the electron wavevector, $v_{\alpha \beta}$ is the electron group velocity, $\epsilon _{n\mathbf{k}}$ is the electron energy, and $I ^{(p)}(\epsilon ,\mu ,T)$ is the material-independent integrand factor:
\begin{equation}
I ^{(p)}(\epsilon ,\mu ,T)=\left ( \epsilon -\mu  \right )^{p}f(\epsilon ,\mu ,T)\left [ 1-f(\epsilon ,\mu ,T) \right ]
\end{equation}
Here, $f(\epsilon ,\mu ,T)$ is the Fermi-Dirac distribution function. The important factor in Eq. (11), electron energy relaxation time $\tau _{n\mathbf{k}}{(\mu ,T)}$ can be defined by considering the electron-phonon coupling effect as follows:
\begin{equation}
\begin{aligned}
&\tau _{(\mu ,T)}^{-1}=\frac{\Omega }{(2\pi )^{2}\hbar}\sum_{m\nu }\int_{BZ}d\mathbf{q}\left | g_{mn\nu }(\mathbf{k},\mathbf{q}) \right |^{2} \\
&\times \{ [ n(\omega _{\nu \mathbf{q}},T)+f(\epsilon _{m\mathbf{k}+\mathbf{q}},\mu ,T) ]\delta (\epsilon _{n\mathbf{k}}+\omega _{\nu \mathbf{q}}-\epsilon _{m\mathbf{k}+\mathbf{q}}) \\
& + [ n(\omega _{\nu \mathbf{q}},T)+1-f(\epsilon _{m\mathbf{k}+\mathbf{q}},\mu ,T) ]\delta (\epsilon _{n\mathbf{k}}-\omega _{\nu \mathbf{q}}-\epsilon _{m\mathbf{k}+\mathbf{q}}) \}
\end{aligned}
\end{equation}
where $\Omega$ is the primitive cell volume, $m$ is the electron band index, $\nu$ is the phonon mode index, $\mathbf{q}$ is the phonon wavevector, $\omega_{\nu}$ is the phonon energy, $n(\omega,T)$ is the Bose-Einstein distribution function, and $\delta$ is the Dirac delta function. Besides, the Eliashberg spectral function related to electron-phonon coupling matrix elements $g_{mn\nu }\mathbf{(k,q)}$ can be defined as\cite{r41}
\begin{equation}
\begin{split}
&\alpha ^{2}F(\omega )=\frac{1}{N(\varepsilon _{F})}\sum_{mn}\sum_{\mathbf{q}\nu }\delta \left( \omega - \omega_{\mathbf{q}\nu }\right ) \sum _{\mathbf{k}}\left |g_{\mathbf{k+q,k}}^{\mathbf{q}\nu ,mn}  \right |^{2} \\
&\times \delta (\varepsilon _{\mathbf{k+q},m}-\varepsilon _{F})\delta (\varepsilon _{\mathbf{k},n}-\varepsilon _{F}).
\end{split}
\end{equation}

However, it is worthy to note that the calculations describing full electron-phonon interaction\cite{r42} as in Eq. (13) are complicated and time-consuming. Hence, in this study, we employed the electron-phonon average approximation\cite{r40} or EPA method which replaces the energy-dependent averages for their momentum-dependent quantities to handle $\tau _{n\mathbf{k}}{(\mu ,T)}$ value. In detail, replacing $\left | g_{mn\nu }(\mathbf{k,q}) \right |^{2} \mapsto g_{\nu }^{2} (\epsilon _{n\mathbf{k}},\epsilon _{m\mathbf{k+q}})$ as the average electron-phonon matrix elements over the directions of $\mathbf{k}$ and $\mathbf{k + q}$ wavevectors, $\omega _{\nu \mathbf{q}} \mapsto \overline{\omega } _{\nu }$ as the average phonon energies over the cells of electron energy grids, and $\rho$ is electron density of states.
\begin{equation}
\begin{split}
&\tau^{-1}(\epsilon ,\mu ,T)=\frac{2\pi \Omega }{g_{s}\hbar}\sum _{\nu } \\
&\{ g_{\nu }^{2}(\epsilon ,\epsilon +\overline{\omega }_{\nu })\left [ n(\overline{\omega }_{\nu },T)+f(\epsilon +\overline{\omega }_{\nu },\mu ,T) \right ] \rho (\epsilon +\overline{\omega }_{\nu }) \\
&+g_{\nu }^{2}(\epsilon ,\epsilon -\overline{\omega }_{\nu })[ n(\overline{\omega }_{\nu },T)+1-f(\epsilon -\overline{\omega }_{\nu },\mu ,T)] \rho (\epsilon +\overline{\omega }_{\nu })\} 
\end{split}
\end{equation}
%
%
%
%
%
%
\section{Results and Discussion}
\subsection{Electronic properties}
Firstly, let us briefly mention to crystal structure issue, which strongly affects electronic structure. The acanthite-like structure of Cu$_{2}$S had been proved as the most stable one at 0 K, theoretically. Under the influence of temperature as shown in FIG. \ref{Fig:pic1}(b), we observed the movement of Cu atoms become chaotic and disordered, this behavior is pretty similar as in the case of \emph{L}-chalc. phase, see FIG. \ref{Fig:pic1}(c). Specifically, S atoms form a nearly immobile sub-lattice with a zig-zag shape while Cu atoms are conductors that moving around disorderly as liquid behavior. To be sure stability of this structure, the phonon dispersion was taken into account (see FIG. \ref{Fig:pic1}(g)). No imaginary frequency appeared in the phonon band structure, revealing that this structure is dynamically stable. It is important to note that, there is the appearance of soft modes at the low-frequency range due to the low crystal symmetry, which leads to the drop-down of some optical modes to acoustic modes. Obviously, this decrease in phonon frequency is usually associated with a certain type of phase transition. Therefore, the assumption that acanthite-like phase can be the pristine phase of Cu$_{2}$S becomes more convincing.\cite{r26} On that basis, it makes sense to investigate about Cu$_{2}$S using this simplified structure.

Minimizing the number of atoms compared to the $L$-chal. phase not only reduces the computational cost but also gives us an insight into the electronic structure, where the states are not overlapping and overly dense. We then started investigating the electronic structure of this model using a GGA functional. FIG. \ref{Fig:pic1}(d) shows that GGA functional completely fails in reproducing the energy gap of Cu$_{2}$S, which is experimentally reported as an indirect bandgap of 1.1 eV.\cite{r24} This is not surprising because GGA functional is known to frequently underestimate the bandgap. Moreover, the strongly correlated nature of transition metal $d$-layer electrons also can be a reason. Hence, the GGA+$U$ method was used to improve the bandgap estimation. At first glance, it can be seen that the Hubbard potential $U$ hardly changes the band edge compared to GGA, the conduction band (CB) is only pushed to the higher energy side than the valence band (VB). As a result, an indirect energy gap of 0.91 eV appears, which is in agreement with the result of previous theoretical work.\cite{r30} For further insight, we also examined band structure using HSE06 functional, which is known to reproduce the bandgap energy of semiconductors better than GGA. The band edge given by HSE06 functional is pretty similar to GGA+$U$ description with VBM located at $\Gamma$--point. Meanwhile, the conduction band minimum (CBM) is distorted along the $\Gamma$--A direction, reproducing a slightly wider indirect bandgap of 0.95 eV compared to the GGA+$U$ result. The second CBM has a slight difference as GGA+$U$ shows that it is located between the Y--C range while HSE06 shows it is located at the Y--point, which may lead to some difference in observing the optical transition states afterward. 
However, it is possible to realize the agreement of the functionals about atomic orbital contributions and their DOSs. The predominance of Cu $d$-orbitals with S $p$-orbitals forms the VB. At the same time, the hybridization between $s$- and $p$-orbitals of Cu, $s$-orbitals of S form the CB. The bandgap estimation of both functionals is still lower than the experimental one, but these values are still reasonable for us to continue investigating other properties of the system.
\begin{figure}[t]
\centerline{\includegraphics{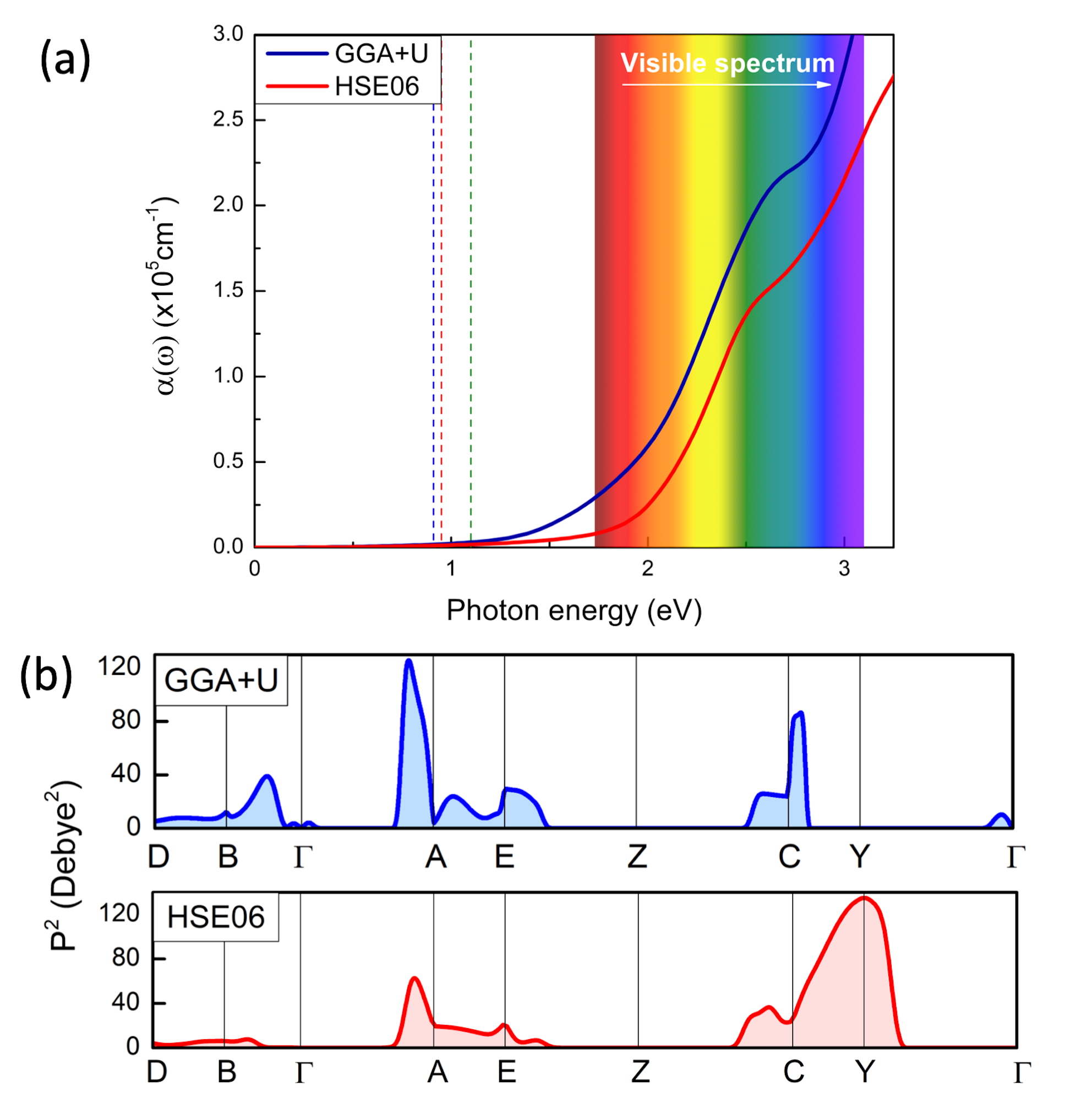}}
\caption{(a) Absorption coefficient as a function of photon energy by GGA+$U$ and HSE06 functionals in the range of the visible light spectrum. The bandgap energy marked by vertical dashed lines (navy: GGA+$U$, red: HSE06, green: experiment data). (b) Transition probability of electric dipole moment between two states of VB-CB along high-symmetry $k$-points.}
\label{Fig:pic2}
\end{figure}

Cu$_{2}$S is known to be a good solar absorber, so it is often fabricated in thin-film form or used as a dopant to increase absorption efficiency.\cite{r43} The high optical absorption coefficient of a material (usually characterized by $\alpha > 10^{4}$ cm$^{-1}$) is important to achieve a good photovoltaic performance. The bandgap of Cu$_{2}$S is in the optimal 1.1--1.7 eV range required for high conversion efficiency as predicted in the Shockley-Queisser limit.\cite{r44} Although the bandgap was underestimated by about 17$\%$ for GGA+$U$ and 13$\%$ for HSE06 functional, both functionals showed a similar trend with a significant increase in $\alpha$ values just about 0.3 eV above the bandgap, as depicted in FIG. \ref{Fig:pic2}(a). The absorption coefficients become characteristic at about 1.8 eV, which is just at the start point of the visible spectrum and increases sharply thereafter. The results calculated by GGA+$U$ functional are greater than those described for HSE06 functional in general, but these values obtained are larger than 10$^{4}$, which is completely competitive with an indirect bandgap absorber as Si.\cite{r45} In addition, the transition probabilities between the two states of the VBM and CBM as shown in FIG. \ref{Fig:pic2}(b) are also revealed in the forbidden or allowed transition state. Accordingly, both functionals agree that the TDM amplitude between VBM and CBM at $\Gamma$--point is zero, indicating that there is no optical absorption between these two states. In contrast, strong optical absorption is observed along the direction of $\Gamma$--A or C--Y. These sites are the lowest points of CB with the distance to VB falls around 2.0 eV. This explains why the magnitude of the absorption coefficient increases sharply from this energy level of the visible spectrum.
\subsection{Intrinsic point defects formation}
\begin{figure}[t]
\centerline{\includegraphics{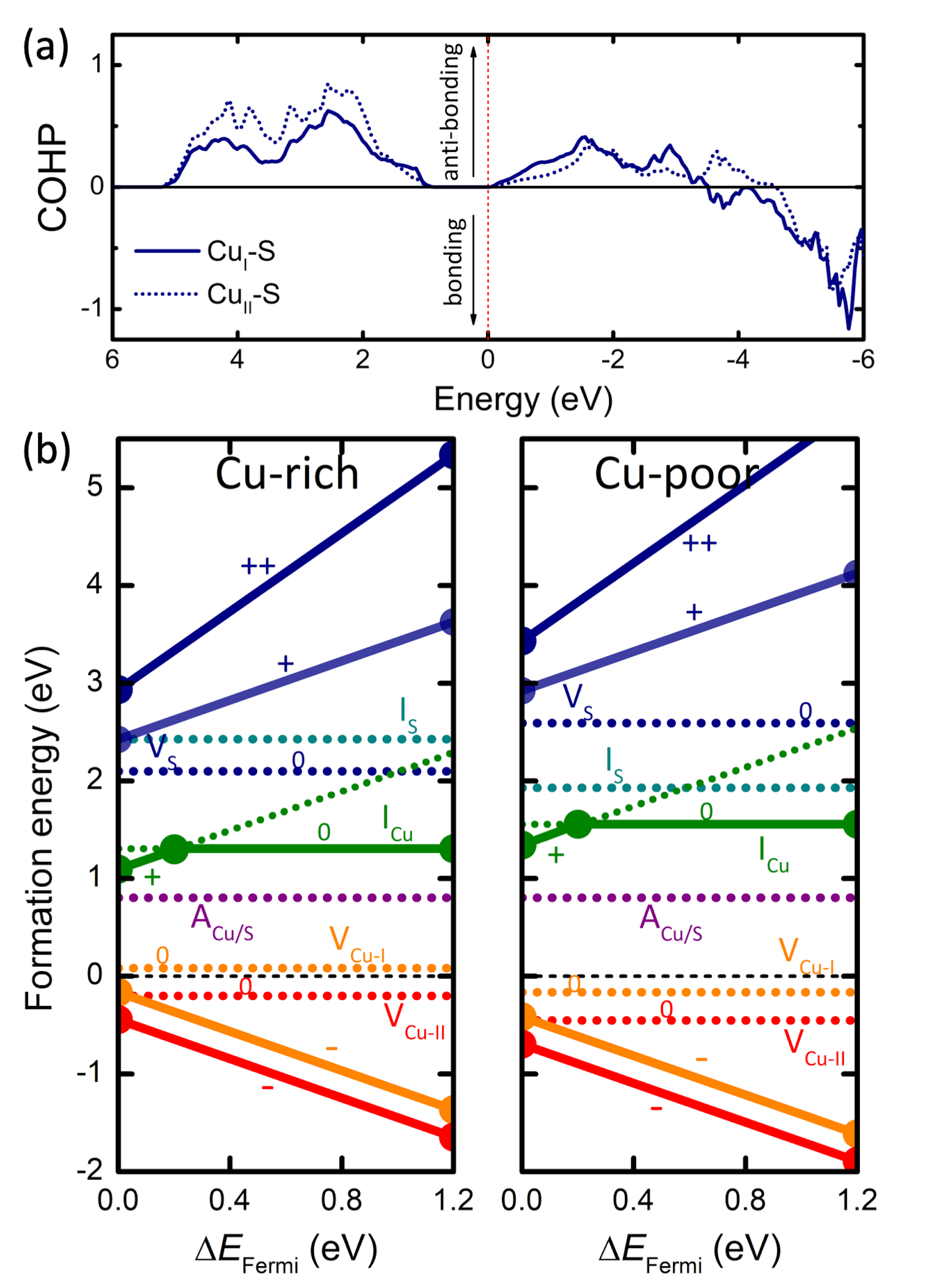}}
\caption{(a) COHP curves of Cu-S bonding where Fermi level (black dashed line) is located at VBM (negative value: bonding nature, positive value: anti-bonding nature), (b) Formation energy of Cu$_{2}$S with different kinds of defects as a function of Fermi level under Cu-rich and Cu-poor conditions. The width of bandgap energy corresponds to horizontal axis values where the VBM point referred to the zero value.}
\label{Fig:pic3}
\end{figure}
\begin{figure}[t]
\centerline{\includegraphics{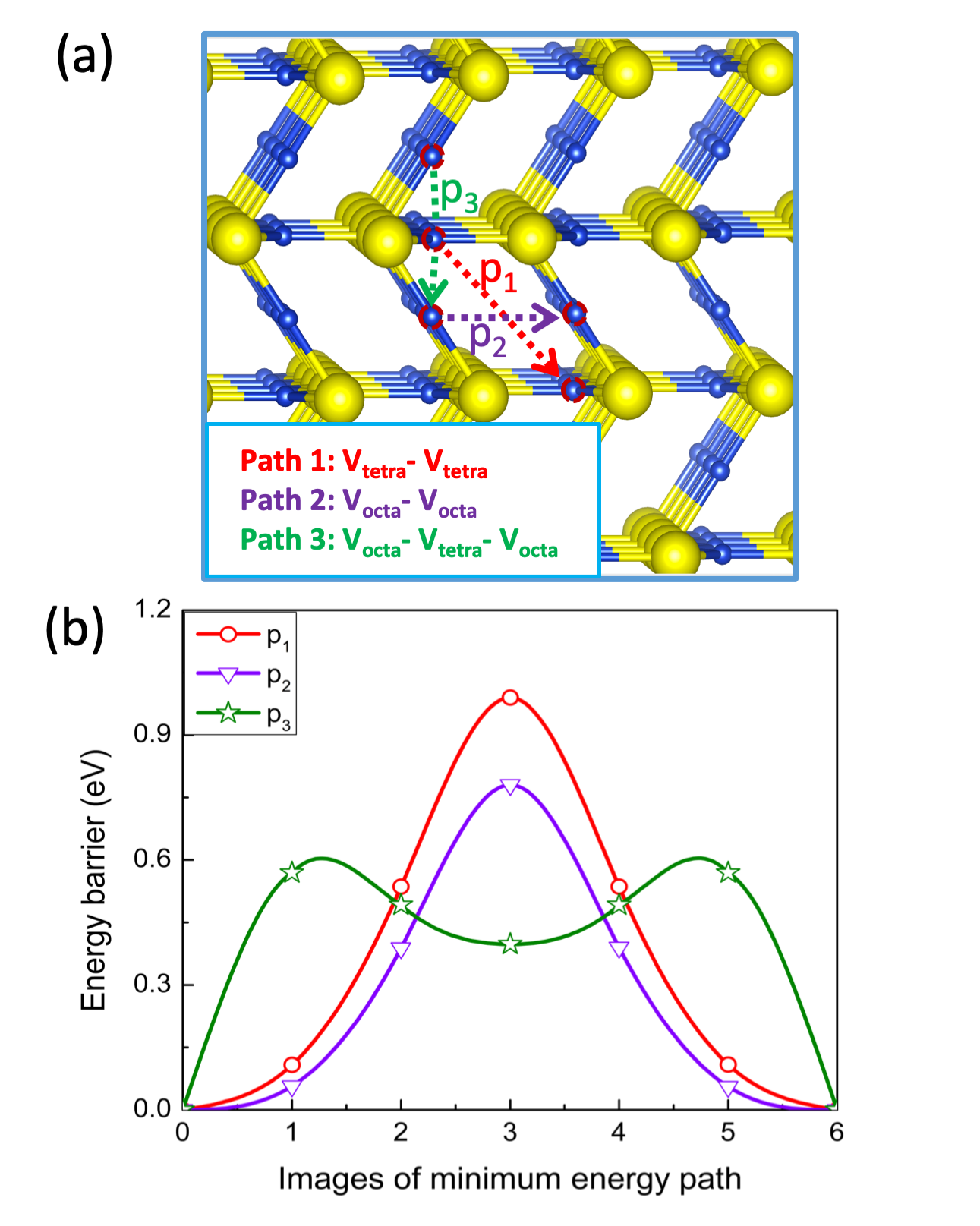}}
\caption{(a) Three possible diffusion pathways of Cu vacancy. (b) MEP of Cu vacancy from T-site to T-site (p$_{1}$), O-site to O-site (p$_{2}$), and O-site to T-site to O-site (p$_{3}$).}
\label{Fig:pic4}
\end{figure}
Here, we mainly consider intrinsic point defects as vacancies, interstitial atoms, and anti-site atoms. Previous studies involving chalcopyrite families such as AgSbTe$_{2}$, CuInSe$_{2}$, CuGaO$_{2}$, CuFeS$_{2}$ showed that they are semiconductors with predominant $p$-type defects.\cite{r46,r47,r48} A common feature that can be observed for this group is the appearance of anti-bonding states between cations and anions below the Fermi level. Consequently, these systems are often structurally unstable and likely to transfer to a more stable form, so that the bonding between cation-anion tends to be easily broken, potentially revealing the formation of defects such as vacancy in the system. FIG. \ref{Fig:pic3}(a) gives us a view of the crystal orbital Hamilton population (COHP) curves between Cu and S in both sites of Cu. Clearly, there is an anti-bonding state between Cu-S at both sites, which is similar to that of the chalcopyrite family. Hence, it leads us to a prediction that $p$-type defects can be the predominant type of defects in the system. To strengthen the above assessment, we further consider the propensity of defects formation based on the energy aspect.

As can be seen in FIG. \ref{Fig:pic3}(b), the formation energy of neutral Cu vacancy at the O-site (i.e.,V$_{\textup{Cu-II}}$ = --0.203 eV at Cu-rich and --0.452 eV at Cu-poor) is smaller than that of the T-site (i.e., V$_{\textup{Cu-I}}$ = 0.083 eV at Cu-rich and --0.165 eV at Cu-poor), meaning O-site is more favored for vacancy formation. Under Cu-poor synthesized conditions, the formation energy of Cu vacancies becomes negative, which means that Cu vacancy is a typical intrinsic defect and extremely easy to form in this system. As a result, the loss of electrons causes the shift of Fermi level to the VB, identified as an acceptor or $p$-type defect. This is consistent with the COHP calculation shown in FIG. \ref{Fig:pic3}(a). Furthermore, the formation energy of Cu vacancy becomes smaller as it accepts an electron and transfers to the charged state --1. The charged state transition between 0/--1 also occurs outside the range of the bandgap energy, indicating that the charged state of the vacancy is preferred over the neutral state. In contrast, the S vacancy formation shows that it is almost unfavorable to form compared to Cu, with an energy of about 2.09--2.59 eV. Therefore, the charged state transitions between 0/+1 and +1/+2 are almost non-existent when the neutral state is assumed to be more stable.

Acting as the $n$-type defect, interstitial Cu atoms become more readily formed under Cu-rich conditions, with a lower amount of energy of about 0.25 eV compared with Cu-poor conditions. In this case, the charged state +1 shows that it is more stable than the neutral state from about 0--0.2 eV above the VBM. Above this energy range, a +1/0 charged state transition occurs, the system changing from one electron-loss to neutral charge at 0.2 eV, above the Fermi level. Similar results are also observed in the case of Ag$_{2}$S,\cite{r49} the appearance of the positively charged interstitial atoms and negatively charged vacancies contributing to the charge neutralization of the system. However, the energetically non-preferred interstitial atoms as in the case of Ag$_{2}$S may be the reason why the vacancy predominance affects conductive behavior as $p$-type of Cu$_{2}$S, which can be seen in the latter part. The interstitial S atom is also considered, although it is evident that this kind of defect is difficult to form in terms of energy, as it requires substantial formation energy of around 2 eV. More energetically preferred than interstitial S atom with a formation energy of about 0.8 eV, the anti-site defect is also a possible intrinsic defect in the system. However, the appearance of this defect almost does not change the conductive properties of Cu$_{2}$S, basically.

Last but not least, being the predominant and most easily formed defect in the system, it is important to look at the diffusion mechanism of Cu vacancy in the system. We simply consider based on MEP, which is more convenient for atoms to migrate. Vacancy migration at the possible pathways that we consider here includes three main ways: from T-site to T-site or V$_{\textup{T}}$--V$_{\textup{T}}$ (p$_{1}$), O-site to O-site or V$_{\textup{O}}$--V$_{\textup{O}}$ (p$_{2}$), and O-site to T-site to O-site or V$_{\textup{O}}$--V$_{\textup{T}}$--V$_{\textup{O}}$ (p$_{3}$), as shown in FIG. \ref{Fig:pic4}(a). The MEP of Cu vacancy as depicted in FIG. \ref{Fig:pic4}(b) express that the maximum energy barrier of p$_{1}$ reaches a value of about 1 eV, which is high enough to challenge the atoms to overcome. Therefore, the migration of the vacancies will of course hardly take place this way because of the high energy barrier. Compared with the p$_{1}$ route, it is easy to see that Cu vacancies will more easily diffuse through the p$_{2}$ route when the maximum point of the energy barrier is only about 0.78 eV. However, the preferred migration pathway of Cu vacancies as predicted could be the p$_{3}$ route. In this way, they just need to overcome a maximum energy barrier of around 0.58 eV from the first O-site to approach the local minimum of the adjacent T-site located around 0.4 eV and then diffuse to the next neighboring O-site, which is easier than the rest of the pathways. A similar migration trend can be found in the case of acanthite Ag$_{2}$S.\cite{r50}
\subsection{Transport properties: electron-phonon coupling}
We know that there are several approaches to determine the transport properties using the Boltzmann theory. Perhaps the most popular and also accessible method is the use of constant relaxation-time approximation (CRTA). However, this strategy often works in systems with good electrical conductors, where the electron energy relaxation time varies very slightly with electron energy, allowing us to regard it as a constant.\cite{r40} Ascertaining electrical conductivity is more challenging, because relaxation time is a direct factor that largely affects the accuracy. As a result, using CRTA for determining this parameter can be a poor approach. Therefore, it would be more prudent to use the EPA method as the main scattering mechanism for our system.

\begin{figure}[t]
\centerline{\includegraphics{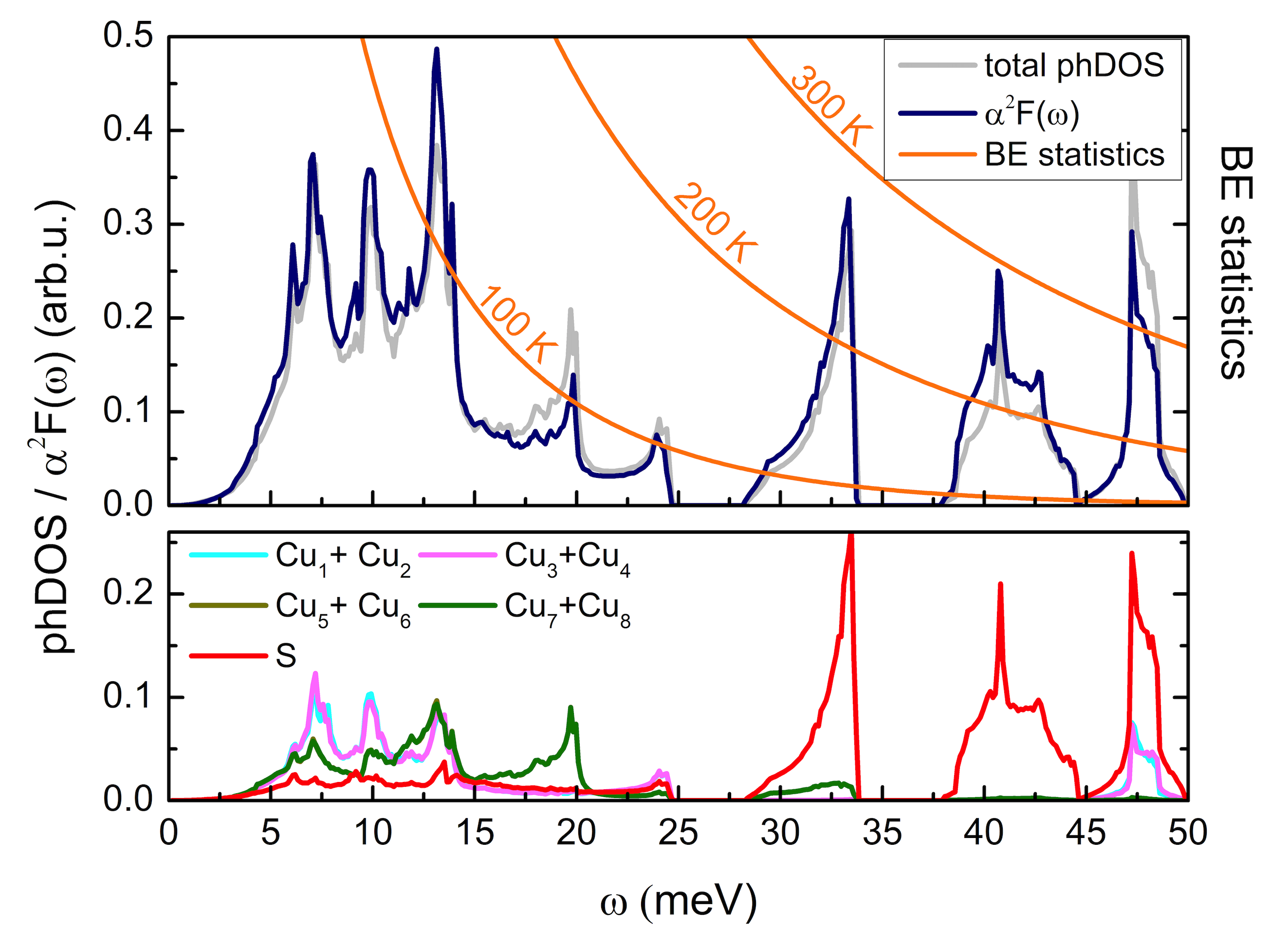}}
\caption{Eliashberg spectral function $\alpha^{2}F(\omega)$ and the total phonon density of states (phDOS) of Cu$_{2}$S (figure above) and the partial phDOS (figure below). The solid orange lines show Bose-Einstein statistics for three lattice temperatures.}
\label{Fig:pic5}
\end{figure}

The Eliashberg spectral function $\alpha^{2}F(\omega)$ is a combination of phonon DOS $F(\omega)$ and the phonon frequency-dependent electron-phonon coupling $\alpha^{2}(\omega)$. By this way, all allowed scattering processes of electrons with phonons of frequency $\omega$ can be observed, as shown in FIG. \ref{Fig:pic5}. Specifically, phonons obey Bose-Einstein statistics in a thermal state at different temperatures. Strong couplings of electrons to phonons can be observed at the ranges of 10--15, 30--35, and 40--45 meV when compared with phonon DOS. In detail, from 10 to 15 meV, the magnitude of Eliashberg function is enhanced due to the strong coupling that occurs on the Cu side (dominated by Cu$_{1}$, Cu$_{2}$, Cu$_{3}$, and Cu$_{4}$). While, the strongest coupling occurs in the high-frequency range of 30--45 meV, mainly caused by scattering on the S side, where phonon occupation getting lower in the thermal state.
\begin{figure}[t]
\centerline{\includegraphics{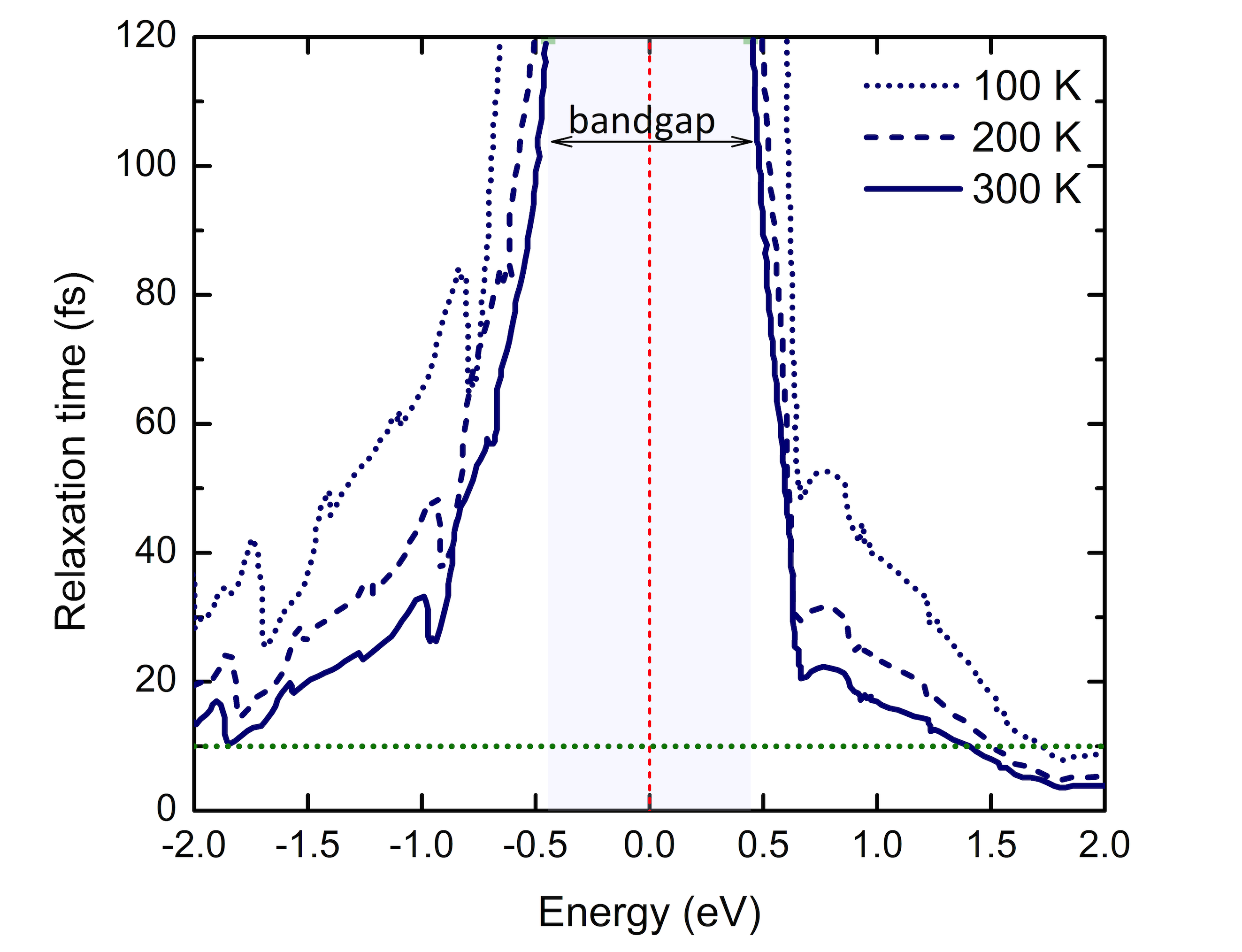}}
\caption{The electron energy relaxation time $\tau$ as a function of the electron energy $\epsilon$ at different temperature ranges. The light navy color bar denotes the bandgap area. Fermi level (red dashed line) is located in the middle of bandgap. Here, $\tau$ = 10 fs (green dashed line) also is used for CRTA method.}
\label{Fig:pic6}
\end{figure}
\begin{figure*}[t] 
\centerline{\includegraphics{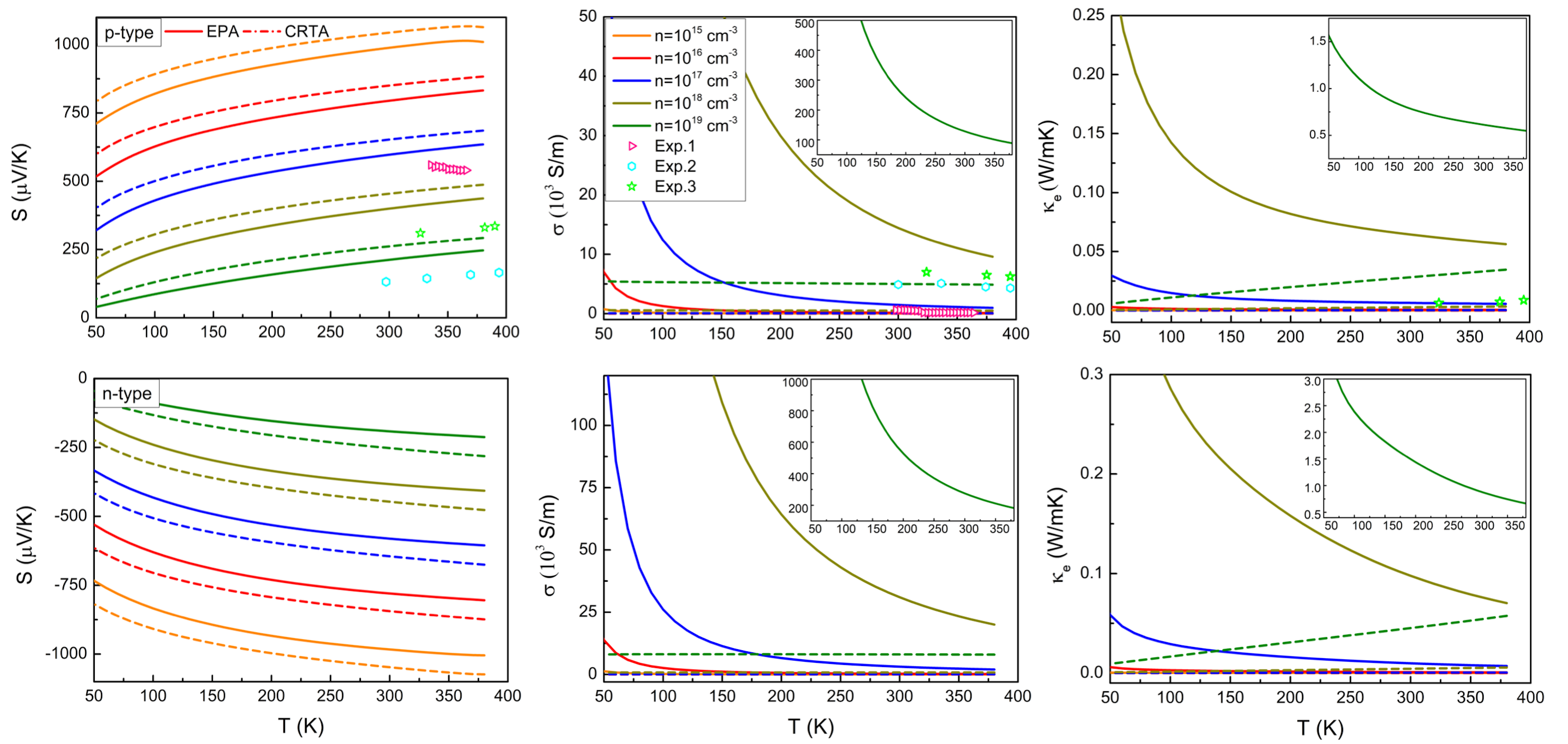}}
\caption{Transport parameters of Cu$_{2}$S acanthite-like phase including Seebeck coefficient (S), electrical conductivity ($\sigma$), and electronic thermal conductivity ($\kappa_{e}$) as a function of temperature at different carrier concentrations. The straight lines indicate theoretical estimation by the EPA method while the dashed lines illustrate results from the CRTA method. Here, data from previous experiment works\cite{r52,r53,r54} is also included for comparison.}
\label{Fig:pic7}
\end{figure*} 
This can be a consequence of the energy transferring between electrons and lattice in non-equilibrium states, resulting in transient non-thermal phonon distributions.\cite{r51} Thereafter, energy dependence of electron relaxation time is calculated and depicted in FIG. \ref{Fig:pic6}. As can be seen, the $\tau$ value shows strong energy dependence and increases sharply near the band edges. This behavior can be explained by the relation:
\begin{equation}
\tau ^{-1}(\epsilon) \sim g^{2}(\epsilon)\rho(\epsilon)
\end{equation}
where $\tau$ is inversely proportional to $\rho$ while electron-phonon matrix element $g^{2}$ shows a weak carrier energy dependency. Because of the strong energy dependence, the contribution of $\tau$ can be considered paramount, especially for conductivity determination. To facilitate comparison, the results from the CRTA method are also taken into account (here, a $\tau = \tau_{\mathrm{const}}$ = 10 fs\cite{r49} is used). Transport parameters are shown in FIG. \ref{Fig:pic7}. Although experimental evidence suggests that the major carrier of Cu$_{2}$S is $p$-type, which agrees with our predictions of the favored-intrinsic defect kind stated previously, we will analyze both $p$- and $n$-type conductive behaviors from a theoretical standpoint below.

Looking at the $p$-type case first, it is clear that EPA method estimates the Seebeck coefficients slightly lower than CRTA method at all carrier concentrations. For instance, at $n$ = 10$^{17}$ cm$^{-3}$ and $T$ = 320 K, CRTA and EPA yield the Seebeck coefficient of 661 and 607 $\mu$VK$^{-1}$. Meanwhile, these values reach 476 and 410 $\mu$VK$^{-1}$ at $n$ = 10$^{18}$ cm$^{-3}$, respectively. Overall, there are no major differences in trend and magnitude for both methods. This is understandable because of the weak dependence of the Seebeck coefficient on $\tau$, which can be eliminated approximately in some cases due to the involvement of $\tau$ in both numerator and denominator, as shown in Eq. (9). It is worthy to note that the Seebeck coefficients reported in previous experimental works show a scattering of several hundred $\mu$VK$^{-1}$ (e.g., at 325 K, the different studies also give different values as around 600 $\mu$VK$^{-1}$,\cite{r52} 310 $\mu$VK$^{-1}$,\cite{r53} and 140 $\mu$VK$^{-1}$\cite{r54}), possibly as a result of differences in sample fabrication methods. This leads to the variation in carrier concentration of these samples, staying around 10$^{17}$--10$^{18}$ cm$^{-3}$. In this case, the EPA gives an estimate that is closer to the experimental value than the CRTA. The difference between the two methods only becomes significant when estimating conductivity.

In CRTA method, $\tau$ behave as a constant, and then, only the change of carrier concentration might not greatly affect conductivity. As can be seen, the $\sigma$ values have almost no significant improvement even though the doping concentration increases. In detail, the change in carrier concentrations at 300 K of 10$^{15}$, 10$^{16}$, 10$^{17}$, 10$^{18}$ cm$^{-3}$ lead to a corresponding change in conductivity of 0.54, 5.1, 50.6, 506 Sm$^{-1}$. That is an improvement of about 10 times in terms of magnitude. Only when increasing the concentration to 10$^{19}$ cm$^{-3}$, $\sigma$ increased dramatically and reproduced the experimental results. Meanwhile, the electron-phonon coupling scattering model shows that $\tau$ is strongly dependent on the energy. This behavior reflects accordingly the significant enhancement of $\sigma$ with each increase in doping concentration. The EPA method estimated electrical conductivity at the same amount of carrier concentrations as the CRTA method but came up with values of 15.6, 147.7, 1465.8, and 14449.1 Sm$^{-1}$, correspondingly. Obviously, the discrepancy in predictions of EPA and CRTA methods under the same conditions is very large. Whether at concentrations 10$^{17}$ or 10$^{18}$ cm$^{-3}$ in the actual measurement,\cite{r52,r53} the EPA method gave a consensus prediction at the same concentration. This means that to reproduce the experimental results, the EPA method gives a reasonable result with the actual concentration while having to increase the concentration in the CRTA method by a minimum of 10 times for a similar prediction.

Similar results are also observed in the case of $\kappa _{e}$. At $n$ = 10$^{18}$ cm$^{-3}$ and 300 K, CRTA method for predictive $\kappa _{e}$ value is 0.0028 Wm$^{-1}$K$^{-1}$.
The significant difference in magnitude by CRTA method only occurred when the doping concentration was up to 10$^{19}$ cm$^{-3}$ (i.e., 0.0283 Wm$^{-1}$K$^{-1}$). Meanwhile, $\kappa _{e}$ values of the same concentration as described by EPA method were 200 times as large (e.g, 0.065 Wm$^{-1}$K$^{-1}$ at 10$^{18}$ cm$^{-3}$ and 0.62 Wm$^{-1}$K$^{-1}$ at 10$^{19}$ cm$^{-3}$). Of course, we can control the $\tau$ value in the CRTA method to best fit the experiment data, but the EPA method plainly shows an advantage in reasonably predicting $\tau$. Thus, taking into account the $\tau$ effect in the conductivity prediction is crucial instead of using it as an input constant. 

A similar trend can be observed when using $n$-type doping. The difference in the Seebeck value of $n$-type versus $p$-type doping is not significant. Although in reality, it is difficult to fabricate the $n$-type conductive sample for Cu$_{2}$S but enhancement of point defects as interstitial Cu atoms can be an approach that helps strengthen the $n$-type carrier. Hence, the conductivity can be greater in magnitude than that of $p$-type doping, theoretically. However, it should also be mentioned that EPA method tends to overestimate electrical and thermal conductivity values at the low-temperature range. This is understandable because the influence of other scattering mechanisms such as impurities, defects, or alloy disorders is ignored in this approximation. Therefore, the addition of these mechanisms in future work could improve the prediction at the low-temperature range.
%
%
%
%
%
%
\section{Conclusions}
In summary, we have performed first-principles calculations incorporating the Boltzmann theory in computing the electronic properties, intrinsic defect formation, and electron-phonon scattering model to estimate the transport properties of Cu$_{2}$S using an acanthite-like model. This theoretical structure not only simplifies the calculation but also confirms the indirect nature of the bandgap as observed before in the experiment. Intrinsic defect formation shows Cu vacancy formation as the most favored defect based on both bonding analysis and energy aspects. Finally, the comparison between CRTA and EPA helps to highlight the suitability of the electron-phonon scattering mechanism in predicting the transport properties of Cu$_{2}$S. For the main purpose, we demonstrate that the acanthite-like model is ideally suitable and can be used for TE material designing purposes related to the low-temperature phase of Cu$_{2}$S.
%
%
%
%
%
%
\section*{Acknowledgement}
This research was supported by JST CREST (Grant No. JP-MJCR18I2). The author H. N. N. acknowledges the financial supports from the Ministry of Education, Culture, Sports, Science, and Technology (MEXT) and Research Grant for Innovative Asia program of Japan International Cooperation Agency (JICA). H. N. N also would like to thank H. B. Tran and T. D. Pham for their valuable help.
%
%
%
%
%
%
\bibliography{}

\end{document}